\documentclass[10pt, a4paper]{article}

\usepackage{lrec-coling2024} 

\usepackage{amsmath}
\usepackage{bm}
\usepackage{enumitem}
\usepackage{xspace}
\usepackage{multirow}
\usepackage{adjustbox}
\usepackage{amsfonts}
\usepackage{graphicx}
\usepackage[normalem]{ulem}
\useunder{\uline}{\ul}{}
\usepackage{url}
\usepackage{balance}

\newcommand{\ours}{MMAPS\xspace}
\newcommand{\hide}[1]{}

\title{\vspace*{.5\baselineskip} \textbf{MMAPS: End-to-End Multi-Grained Multi-Modal Attribute-Aware Product Summarization}}

\name{Tao Chen\textsuperscript{\rm 1}, Ze Lin\textsuperscript{\rm 1}, Hui Li\textsuperscript{\rm 1 *}\thanks{* Corresponding Author.}, Jiayi Ji\textsuperscript{\rm 1}, Yiyi Zhou\textsuperscript{\rm 1}, Guanbin Li\textsuperscript{\rm 2}, Rongrong Ji\textsuperscript{\rm 1}} 

\address{\textsuperscript{\rm 1}Key Laboratory of Multimedia Trusted Perception and Efficient Computing, Ministry of Education of China\\Xiamen University\\ \textsuperscript{\rm 2}School of Computer Science and Engineering, Sun Yat-sen University \\
         \{taochen, linze4014\}@stu.xmu.edu.cn, \{hui, zhouyiyi, rrji\}@xmu.edu.cn\\
         jjyxmu@gmail.com, liguanbin@mail.sysu.edu.cn\\}

\abstract{
Given the long textual product information and the product image, Multi-modal Product Summarization (MPS) aims to increase customers' desire to purchase by highlighting product characteristics with a short textual summary. Existing MPS methods can produce promising results. Nevertheless, they still 1) lack end-to-end product summarization, 2) lack multi-grained multi-modal modeling, and 3) lack multi-modal attribute modeling. To improve MPS, we propose an end-to-end multi-grained multi-modal attribute-aware product summarization method (MMAPS) for generating high-quality product summaries in e-commerce. MMAPS jointly models product attributes and generates product summaries. We design several multi-grained multi-modal tasks to better guide the multi-modal learning of MMAPS. Furthermore, we model product attributes based on both text and image modalities so that multi-modal product characteristics can be manifested in the generated summaries. Extensive experiments on a real large-scale Chinese e-commence dataset demonstrate that our model outperforms state-of-the-art product summarization methods w.r.t. several summarization metrics. Our code is publicly available at: \url{https://github.com/KDEGroup/MMAPS}.
 \\ \newline  }

\begin{document}

\pagenumbering{gobble}
\maketitleabstract

\section{Introduction}
\label{sec:intro}

\begin{figure}[t]  
    \centering  
    \includegraphics[width=0.5\textwidth]{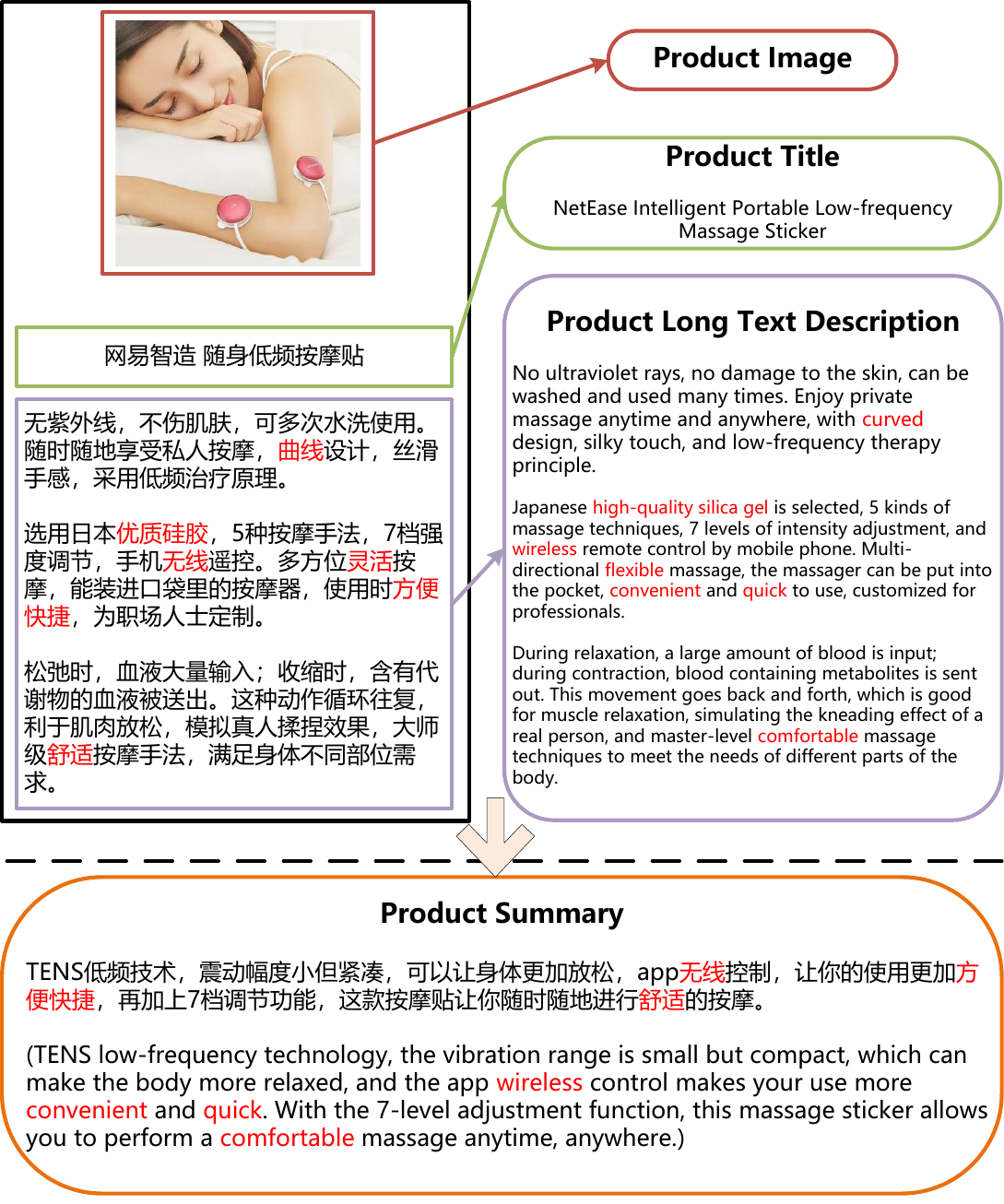}  
    \caption{An example of a product in the CEPSUM dataset~\cite{LiYXWHZ20}. Product attributes are shown in red.}
    \label{fig:dataexample}  
\end{figure}

With the development of the Internet, online shopping has become an integral part of people's daily life. 
Unlike brick-and-mortar stores, where customers can interact face-to-face with salespeople, people mainly learn about products through textual and pictorial descriptions in online stores. 
A product is typically described with a product title, one or a few product images and a long product description in the online store.
For instance, Fig.~\ref{fig:dataexample} provides an example of a portable massage sticker sold in a Chinese online store.

However, long product descriptions increase the cognitive load and hurt the shopping experience. 
Hence, informative product summaries are critical for online stores to provide a better shopping experience and boost product sales. 
As new products emerge rapidly, manual summarization becomes cost prohibitive, leaving alone that it requires a certain level of expertise to write accurate and attractive product summaries. 
To overcome this problem, much effort has been devoted to designing product summarization methods~\cite{ChenLZYZ019,LiYXWHZ20,SongJLZCN22} that automatically summarizes 
product information to highlight product's characteristics and advantages.

Early methods~\cite{WangHLCL17, abs-1807-08000, ChenLZYZ019, DaultaniNC19} mainly focus on leveraging product's textual information, such as descriptions and attributes. 
However, they ignore the product's visual modality which can provide rich information for product summarization.
Recently, some works~\cite{ZhangZLWPGY19,LiYXWHZ20,SongJLZCN22} consider multi-modal product information in product summarization, i.e., Multi-modal Product Summarization (MPS).
As illustrated in Fig.~\ref{fig:dataexample}, both the text modality and the visual modality are presented on product's web page and they convey cues regarding key product characteristics which are helpful information for generating eye-catching product summaries. 
Existing MPS methods have achieved promising performance.
Nevertheless, they still suffer from several problems:
\begin{itemize}[leftmargin=10pt,topsep=1pt,itemsep=0.3pt]

    \item \textbf{P1: Lack end-to-end product summarization.} State-of-the-art MPS approaches~\cite{LiYXWHZ20,SongJLZCN22} treat product attribute modeling and product summarization generation as two separate phases and train them independently, making it difficult to train and tune the complete process.
    Moreover, the error from the two phases is accumulated, negatively affecting the summarization.

    \item \textbf{P2: Lack multi-grained multi-modal modeling.} Existing MPS methods~\cite{LiYXWHZ20,SongJLZCN22} fuse the global information (e.g., sentence-level representation and image-level representation) from different modalities, i.e., coarse-grained fusion.
    They neglect the importance of fine-grained multi-modal modeling (e.g., token-level and region-level representations alignment and fine-grained multi-modal product attribute modeling), which can improve the quality of generated product summaries.

    \item \textbf{P3: Lack multi-modal attribute modeling.} Product attributes describe key product characteristics (e.g., design and functionality) which help MPS models better understand products. However, existing works only model  attributes from one modality, leading to inferior results. For instance, V2P~\cite{SongJLZCN22} models product attributes by extracting features from product images, while MMPG~\cite{LiYXWHZ20} considers attribute features from text.
\end{itemize}

To address the above issues, we propose an end-to-end \underline{M}ulti-Grained \underline{M}ulti-modal \underline{A}ttribute-aware \underline{P}roduct \underline{S}ummarization method (\ours).
Our contributions are summarized as follows:
\begin{itemize}[leftmargin=10pt,topsep=1pt,itemsep=0.3pt]

\item To deal with P1, we design an end-to-end multi-modal product summarization method \ours, which jointly models product attributes and summary generation. \ours can attend to the product characteristics from multiple modalities, helping generate coherent product summaries. The end-to-end learning process also reduces the difficulty of training and tuning.

\item To remedy P2, we propose several multi-grained multi-modal tasks to guide \ours. We design a coarse-grained dual-encoder contrastive learning task to align cross-modal coarse-grained information coarsely. Additionally, we design the fine-grained multi-modal alignment task and the fine-grained product attribute prediction task to endow \ours with the ability to capture fine-grained cross-modal product information.

\item To handle P3, in the fine-grained product attribute prediction task, \ours models product attributes based on both text and image modality, which helps \ours understands multi-modal product characteristics and guides \ours to pay more attention to the significant features when generating product summaries.

\item We conduct extensive experiments on a large-scale Chinese e-commerce dataset. Experimental results demonstrate that \ours outperforms state-of-the-art product summarization methods.

\end{itemize}

\section{Related Work}

\subsection{Product Summarization}

E-Commerce product summarization aims to generate text summaries that provide customers with the most valuable information about the product, increasing their desire to purchase the product.  

Traditional works only take the long textual descriptions of the products as input. 
For example, Yuan et al.~\cite{YuanLXWHZ20} propose to use a dual-copy mechanism to generate faithful product summaries, which can selectively copy tokens from both product descriptions and attributes. 
EPCCG~\cite{0002Z0XLW22} utilizes Transformer~\cite{VaswaniSPUJGKP17} to summarize controllable product copywriting from product title, attributes, and OCR text in different aspects.

Despite their promising performance, earlier methods do not incorporate the products' visual signal in the text generation.
The visual modality can help customers discriminate essential product characteristics and thus improve the quality of the generated summaries. 

Recently, a few works resort to multi-modal product summarization which considers multi-modal information.
For example, MMPG~\cite{LiYXWHZ20} introduce a multi-modal pointer-generator network to generate an aspect-aware textual summary for Chinese e-commerce products by integrating textual and visual product information. 

V2P~\cite{SongJLZCN22} unifies the heterogeneous multi-modal data in the same embedding space by converting the vision modality into semantic attribute prompts.

\subsection{Multi-Modal Self-Supervised Learning}

Self-supervised learning has been widely used in pre-trained models~\cite{LiuZHMWZT23}.
In multi-modal learning, many works~\cite{LiZZZWJWW21, ShengPTL21, LinSY020} 
have exploited contrastive learning, a type of self-supervised learning, to model the correspondence among different modalities. 
For instance, Oscar~\cite{Li0LZHZWH0WCG20} learns cross-modal representations by predicting whether the image-text pair contains the original image representation or any polluted one. 
ALBEF~\cite{LiSGJXH21} learns a similarity function such that parallel image-text pairs have higher similarity scores.
Based on the learned function, 
ALBEF enforces the representations of an image and a text in a pair close to each other. 
AVTS~\cite{KorbarTT18} leverage the consistency of videos and audios to train the deep encoder. 
MMV~\cite{AlayracRSARFSDZ20} further considers the consistency among video, audio and text.

\begin{figure*}[t]
    \begin{center}
        \includegraphics[width=0.93\textwidth]{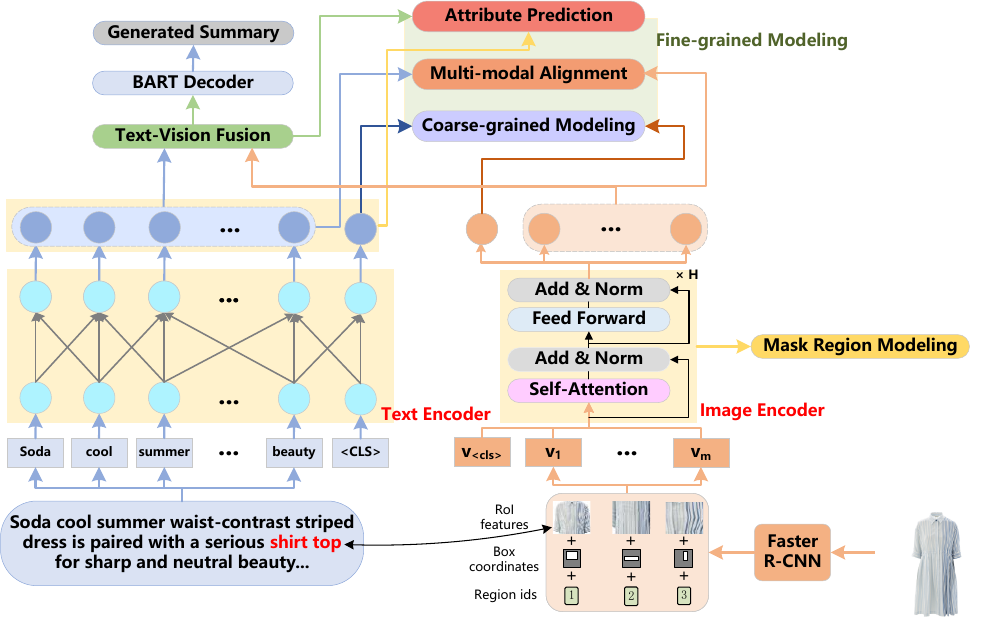}
        \caption{Overview of \ours. Chinese product information has been translated into English.}
        \label{fig:overview}
    \end{center}
\end{figure*}

\section{Our Method \ours}

In this section, we illustrate the details of our proposed \ours.
Fig.~\ref{fig:overview} provides an overview of \ours.
\ours takes the product information $X$, which contains the title and the long description of a product, and the product image $I$ as input, and it generates a product summary that is much shorter than the product description.

\subsection{Architecture Design}

The architecture of \ours consists of four components. 
We directly use the design of BART decoder as the decoder in \ours to decode summaries.
Text encoding and image encoding components are used for encoding textual features and visual features, respectively. 
The text-image fusion component is designed for fusing multi-modal representations that are prepared for the product summarization. 
In the following, we illustrate the details of text encoding, image encoding and text-image fusion component.

\subsubsection{Text Encoding}

The product information $X$ is tokenized and a special token [CLS] is appended to the token sequence in order to represent the overall, sequence-level information after text encoding.
Then, we use the pre-trained BART encoder as the text encoder in \ours and the token sequence is fed into it.
The output encoded token representation sequence is denoted as $\mathbf{Z}=\left\{\mathbf{z}_{1},\mathbf{z}_{2},\cdots,\mathbf{z}_{L},\mathbf{z}_{\text{<cls>}}\right\}$ where $L$ is the length of the sequence, $\mathbf{z}_{i}$ is the encoded representation of $i$-th token, and $\mathbf{z}_{\text{<cls>}}$ is the representation of [CLS].
$\mathbf{z}_{\text{<cls>}}$ is inferred by all tokens in the sequence and it can be regarded as the sequence-level representation.

\subsubsection{Image Encoding}
\label{sec:ie}

We use the Faster R-CNN~\cite{RenHGS15} pre-trained on Visual Genome~\cite{KrishnaZGJHKCKL17} to extract region representations.
Specifically, we feed the input image $I$ to the pre-trained Faster R-CNN and extract all the detected objects.
We only retain $m$ objects with the highest confidence. 
To keep the spatial information of the image, each image region $i$ is encoded as the sum of three types of features~\cite{ChoLTB21}: 
\begin{enumerate}[leftmargin=14pt,topsep=1pt,itemsep=0.3pt]
\item RoI (Region of Interest) object feature $\mathbf{v}_{i}$; 

\item RoI bounding box coordinate feature $\mathbf{e}_{i}^{\text{box}}$, which is encoded with a linear layer; 

\item Region id feature $\mathbf{e}_{i}^{\text{reg}}$, which is encoded by the embedding layer in the text encoder. 
\end{enumerate}
The final region representation $\mathbf{o}_{i}$ of $i$ can be obtained as follows:
\begin{equation}
    \label{eq:region}
    \small
    \begin{aligned}
        &\mathbf{v}_{i}, \mathbf{c}_{i}, \mathbf{r}_{i} \in \text{Faster R-CNN}(I) \\
        &\mathbf{e}_{i}^{\text{box}} = [\mathbf{c}_{i},\mathbf{s}_{i}]\mathbf{W}_{e} + \mathbf{B}_{e},\,\,\,\,\mathbf{o}_{i} = \mathbf{v}_{i} + \mathbf{e}_{i}^{\text{box}} + \mathbf{e}_{i}^{\text{reg}} 
    \end{aligned}
\end{equation}
where $\mathbf{c}_{i}$ is a 4-dimensional normalized position vector indicating the coordinates of the top-left and bottom-right corners, 
$\mathbf{s}_{i}\in \mathbb{R}^{1}$ indicates the corresponding area of $\mathbf{c}_{i}$, $\mathbf{r}_{i}$ is the class distribution of region $i$ which is later used in the Masked Region Modeling task introduced in Sec.~\ref{sec:vs}, $[\cdot,\cdot]$ represents the concatenation operation,
and $\mathbf{W}_{e}$ and $\mathbf{B}_{e}$ are learnable parameters.
$\text{Faster R-CNN}(I)$ outputs $m$ triples $\{\mathbf{v}, \mathbf{c}, \mathbf{r}\}$ and each of them corresponds to an image region.
If an image has less than $m$ detected regions, we use vectors with all zeros as $\mathbf{o}$ to pad region representations to have $m$ region representations for that image.

$\mathbf{o}$ is only position-based region representation.
To further model the context of the region, we subsequently feed $\mathbf{o}$ into a Transformer-based unimodal encoder.
For an image, we construct a region representation sequence with its $m$ region representations: $\mathbf{O}=\left\{\mathbf{o}_{\text{<cls>}},\mathbf{o}_{1},\mathbf{o}_{2},\cdots,\mathbf{o}_{m}\right\}$.
Similar to the [CLS] token used in text encoding, we prepend a learnable parameter $\mathbf{o}_{\text{<cls>}}$ to $\mathbf{O}$ to represent the global representation of the image. 
The image encoder consists of $H$ stacked layers and each layer includes two sub-layers: $1)$ Multi-Head Attention (MHA) and $2)$ a position-wise Feed-Forward Network (FFN). Finally, we obtain the output visual features $ \mathbf{G}=\left\{\mathbf{g}_{\text{<cls>}},\mathbf{g}_{1},\mathbf{g}_{2},...,\mathbf{g}_{m}\right\} $:
\begin{equation}
    \small
    \mathbf{S}^{h}_{V} = \text{MHA}(\mathbf{U}^{h-1}_{V}) + \mathbf{U}^{h-1}_{V},\,\,\,\,\mathbf{U}^{h}_{V} = \text{FFN}(\mathbf{S}^{h}_{V}) + \mathbf{S}^{h}_{V} 
    \label{eq:visual}
\end{equation}
where $\mathbf{U}^{0}_{V}$ is the input region representations $\mathbf{O}$.

\subsubsection{Text-Image Fusion}

\ours contains a multi-modal fusion component. 
It receives the encoded text features $\mathbf{Z}$ and image features $\mathbf{G}$ from the text and image encoding component 
and outputs fused, multi-modal features containing related information derived from textual and visual modality, aiming to complement the information of textual modality while potentially removing noisy information. 

The process of the multi-modal fusion is shown as follows:
\begin{equation}
    \small
    \begin{split}
        &\mathbf{Q} =\mathbf{ZW}_{q},\,\,\,\,\mathbf{K}=\mathbf{GW}_{k},\,\,\,\,\mathbf{V}=\mathbf{GW}_{v} \\
        &\mathbf{C} = \text{CMA}(\mathbf{Q},\mathbf{K},\mathbf{V}),\,\,\,\,\mathbf{F} = \sigma([\mathbf{Z},\mathbf{C}]\mathbf{W}_{f} + \mathbf{B}_{f}) \\
        &\mathbf{Z}^{'} = [\mathbf{Z},\mathbf{F} \otimes \mathbf{C}]\mathbf{W}_{z^{'}} + \mathbf{B}_{z^{'}} \\
    \end{split}
\end{equation}
where $\text{CMA}(\cdot)$ indicate the cross-modality attention layer using the Multi-Head Attention mechanism (i.e., $\text{MHA}(\cdot)$ in Eq.~\ref{eq:visual} using different input as query, key and value), $\sigma(\cdot)$ refers to the Sigmoid function, $\mathbf{W}_{*}$ and $\mathbf{B}_{*}$ are learnable parameters, and $\otimes$ is element-wise vector multiplication.
Specifically, the textual features $\mathbf{Z}$ are linearly projected into queries $\mathbf{Q}$ and the visual features $\mathbf{G}$ are linearly mapped to key-value pairs $\mathbf{K}$ and $\mathbf{V}$. 
Next, to retain the pre-trained text features from BART and overcome the noise brought by the visual modality, we apply a forget gate $\mathbf{F}$.
Finally, we concatenate the textual features $\mathbf{Z}$ and the result of $\mathbf{F} \otimes \mathbf{C}$ to generate the multi-modal features $\mathbf{Z}^{'}$.

\subsection{Multi-Modal Multi-task Learning}

We design several tasks to guide the multi-modal learning of \ours:

\subsubsection{Product Summarization}

To supervise the product summarization task, given input text $X$ and image $I$, we fulfill the output-level supervision by minimizing the negative log-likelihood:
\begin{equation}
    \small
    \mathcal{L}_\text{PS} = -\sum_{y\in Y}\sum_{t=1}^{\left|y\right|} log\big(p(y_{t}|I,X,y_{1},...,y_{t-1})\big), 
\end{equation}
where $y_t$ is the $t$-th token in the ground truth summary $y$, $\left|y\right|$ is the number of tokens in $y$ and $Y$ is the ground truth summary set.

\subsubsection{Masked Region Modeling}
\label{sec:vs}

We adopt the Masked Region Modeling task to improve the performance of the image encoder
For input images, \ours samples $S$ image regions with a probability of 15\% and masks the sampled regions.
Then, we train the model to predict the class of the masked regions.
For the $s$-th masked region, we use $\mathbf{g}_{s}$ and $\mathbf{r}_{s}$ to denote the visual feature output by the image encoder and the class distribution detected by Faster R-CNN (i.e., $\mathbf{r}$ in Eq.~\ref{eq:region}), respectively.
\ours minimizes the KL divergence of the predicted class distribution and $\mathbf{r}_{s}$:
\begin{equation}
    \small
    \mathcal{L}_\text{MRM} = \sum_{s=1}^{S} D_{KL}\big(\mathbf{r}_{s} \| \text{MLP}(\mathbf{g}_{s})\big), 
\end{equation}
where $\text{MLP}(\cdot)$ is a two-layer perception for classification.

\subsubsection{Multi-grained Multi-modal Modeling}
As product summarization involves two correlated modalities, modeling the correlation between the text modality and the image modality will help \ours generate expressive text that corresponds to the product image.
Moreover, both coarse-grained information (sentence-level representation and image-level representation) and fine-grained information (token-level representation and region-level representation) from the two modalities are related to depicting the product and their cross-modal correlation affects the quality of the generated product summaries.
Consequently, we design coarse-grained modeling and fine-grained modeling to help \ours better capture the cross-modal correlation.

\noindent\textbf{Coarse-grained Multi-modal Modeling.}
We design a coarse-grained dual-encoder contrastive learning task to jointly optimize the image encoder and the text encoder by contrasting the text-image pairs against others in the same batch. 
Specifically, for each image-text pair (positive pair) in a batch of $B$ image-text pairs, we use its image and all the text in the remaining $B-1$ pairs to form $B-1$ negative image-text pairs.
Then, the image encoder and the text encoder are trained to maximize the similarity between the image-level representation ($\mathbf{g}_\text{<cls>}$) and the sentence-level representation ($\mathbf{z}_\text{<cls>}$) for a positive pair and minimize the similarity of representations corresponding to the $B(B-1)$ negative pairs.
The loss of the contrastive learning is shown as follows:
\begin{equation}
\label{eq:cl}
\small
\begin{aligned}
    &\mathbf{g}_\text{norm}^{i} =\mathbf{W}_{g}\mathbf{g}_\text{<cls>}^{i} \big/ \left\| \mathbf{W}_{g}\mathbf{g}_\text{<cls>}^{i} \right\|_{2} \\
    &\mathbf{z}_\text{norm}^{j} =\mathbf{W}_{z}\mathbf{z}_\text{<cls>}^{j} \big/ \left\| \mathbf{W}_{z}\mathbf{z}_\text{<cls>}^{j} \right\|_{2} \\
    &\mathcal{L}_\text{i2t} = -\frac{1}{B}\sum_{i}^{B}log\frac{exp(\mathbf{g}_\text{norm}^{i\top}\mathbf{z}_\text{norm}^{i} \big/ \tau)}{\sum_{j=1}^{B}exp(\mathbf{g}_\text{norm}^{i\top}\mathbf{z}_\text{norm}^{j} \big/ \tau)} \\
    &\mathcal{L}_\text{t2i} = -\frac{1}{B}\sum_{i}^{B}log\frac{exp(\mathbf{z}_\text{norm}^{i\top}\mathbf{g}_\text{norm}^{i} \big/ \tau)}{\sum_{j=1}^{B}exp(\mathbf{z}_\text{norm}^{i\top}\mathbf{g}_\text{norm}^{j} \big/ \tau)} \\
    &\mathcal{L}_\text{CMM} = \mathcal{L}_\text{i2t} + \mathcal{L}_\text{t2i} \\
\end{aligned}
\end{equation}
where $\mathbf{g}_\text{norm}^{i}$ and $\mathbf{z}_\text{norm}^{j}$ are normalized representations of the image in the $i$-th pair and the text in the $j$-th pair, respectively.
$\mathbf{W}_{g}$ and $\mathbf{W}_{z}$ are weights and $\tau$ is a learnable temperature parameter.

\noindent\textbf{Fine-grained Multi-modal Modeling.} 
The coarse-grained multi-modal modeling improves the quality of sequence-level representations and image-level representations via aligning global semantic information.
Besides, product attributes, which depict the key product characteristics that customers are most concerned with, also require fine-grained multi-modal modeling to better model attributes from two modalities.
Therefore, we further propose two fine-grained multi-modal modeling tasks to enhance \ours.
\begin{enumerate}[leftmargin=14pt,topsep=1pt,itemsep=0.3pt]
    \item \textbf{Fine-grained Multi-modal Alignment.}
    We design a fine-grained multi-modal alignment task to train \ours to model the semantic correlation between the text modality and the image modality by inspecting individual regions and tokens. 
    This way, \ours can capture the salient information that appears in both modalities. 
    To do this, we use Hausdorff distance, which can measure the similarity between two sequences of different features, 
    to align textual and visual features:
    \begin{equation}
    \label{eq:hd}
    \small
    \begin{split}
        &d(\mathbf{G},\mathbf{Z}) = \displaystyle \max_{i} \displaystyle \min_{j}\left\|\frac{\text{MLP}(\mathbf{g}_{i})}{\left\| \text{MLP}(\mathbf{g}_{i}) \right\|_{2}} - \frac{\mathbf{z}_{j}}{\left\| \mathbf{z}_{j} \right\|_{2}}\right\|_{2} \\
        &d(\mathbf{Z},\mathbf{G}) = \displaystyle \max_{j} \displaystyle \min_{i}\left\|\frac{\text{MLP}(\mathbf{g}_{i})}{\left\| \text{MLP}(\mathbf{g}_{i}) \right\|_{2}} - \frac{\mathbf{z}_{j}}{\left\| \mathbf{z}_{j} \right\|_{2}}\right\|_{2} \\
        &d_{H}(\mathbf{G},\mathbf{Z}) = \displaystyle \max\big\{ d(\mathbf{G},\mathbf{Z}), d(\mathbf{Z},\mathbf{G})\big\}\\
        &\mathcal{L}_\text{HD} = d_{H}^{2} \\
    \end{split}
    \end{equation}
    In Eq.~\ref{eq:hd}, we use a two-layer perception $\text{MLP}(\cdot)$ to map the encoded visual features into the text representation space. 
    The above Hausdorff Loss can capture the local boundary information and enforce multi-modal alignment, making features in different modalities that correspond to the same characteristics getting closer in the common semantic space.

    \item \textbf{Fine-grained Multi-Modal Product Attribute Prediction.} 
    Each product in e-commerce platform is typically described by some product attributes.
    For instance, Fig.~\ref{fig:overview} shows that the product attribute ``shirt top'' is manifested in both the text and the image.
    Including the descriptions of product attributes in the generated text improves the quality of product summaries and makes the summaries eye-catching.
    Moreover, they help customers quickly understand and distinguish the bright spots of different products.
    Therefore, we design a multi-modal product attribute prediction task, which trains \ours to predict product attributes, to endow \ours with the ability of understanding multi-modal product characteristics and guide \ours to pay more attention to the significant features when generating product summaries.
    For a product in the dataset, we assign a multi-hot vector as the ground-truth attribute vector, denoted as $\mathbf{y}^{a} = (\mathbf{y}^{a}_{1}, \cdots, \mathbf{y}^{a}_{N})$, where $\mathbf{y}^{a}_{l}=1$ denotes the product has the $l$-th attribute, $\mathbf{y}^{a}_{l}=0$ otherwise. $N$ is the total number of the attributes. 
    To predict product attributes, we feed the text representation $\mathbf{Z}$ and the multi-modal representation $\mathbf{Z}^{'}$ into a feed-forward layer to predict the attribute vector:
    \begin{equation}
        \small
        \begin{aligned}
        \hat{\mathbf{y}}^{a} = &\sigma\big(\mathbf{W}_{y}^{(1)}(\mathbf{W}_{y}^{(2)}\sum_{i=1}^{L}\mathbf{z}_{i}^{'} + \mathbf{W}_{y}^{(3)}\sum_{j=1}^{L}\mathbf{z}_{j} + \mathbf{W}_{y}^{(4)}\mathbf{z}_{<cls>}) \\
        & + \mathbf{B}_{y}\big), 
        \end{aligned}
    \end{equation}
    where $\mathbf{W}$ and $\mathbf{B}$ are trainable weights, and $\hat{\mathbf{y}}^{a} $ indicates the predicted attributes.    
    We use a binary cross entropy loss for the multi-modal product attribute prediction task:
    \begin{equation}
        \small
        \mathcal{L}_\text{ATT} = \displaystyle \min_{\Theta_{a}} - \sum^{B} \big(\mathbf{y}^{a} ln(\hat{\mathbf{y}}^{a}) + (\mathbf{1} - \mathbf{y}^{a}) ln(\mathbf{1} - \hat{\mathbf{y}}^{a})\big), 
    \end{equation}
    where $\mathbf{1} \in \mathbb{R}^{N}$ is a vector with all elements being one, 
    and $\Theta_{a}$ stands for corresponding to-be-learned parameters.

\end{enumerate}

\begin{table}[]
    \resizebox{\columnwidth}{!}{
    \begin{tabular}{cccc}
    \hline
    Category              & Home Appliances & Clothing & Cases \& Bags \\ \hline
    \#Train Sample        & 437,646         & 790,297  & 97,510        \\
    \#Valid Sample        & 10,000          & 10,000   & 5,000         \\
    \#Test Sample         & 10,000          & 10,000   & 5,000         \\
    Avg. Length of Input  & 335             & 286      & 299           \\
    Avg. Length of Output & 79              & 78       & 79            \\ \hline
    \end{tabular}
    }
    \caption{Statistics of data. The unit of the average length of input/output is one Chinese character.}
    \label{data_static}
\end{table}

\subsection{Putting All Together}

In summary, training \ours involves optimizing four parts and the overall objective is defined as follows:
\begin{equation}
    \small
    \mathcal{L} = \mathcal{L}_\text{PS} + \lambda_{1}\mathcal{L}_\text{MRM} + \lambda_{2}\mathcal{L}_\text{CMM} + \lambda_{3}\mathcal{L}_\text{FMM},
    \label{eq:eq-all} 
\end{equation}
where $\mathcal{L}_\text{FMM}=\mathcal{L}_\text{HD}+\mathcal{L}_\text{ATT}$ and $\lambda_{*}$ are pre-defined loss weights.

\section{Experiment}
\label{sec:exp}

In this section, we report and analyze the experimental results in order 
to answer the following research questions:
\begin{itemize}[leftmargin=10pt,topsep=1pt,itemsep=0.3pt]

\item \textbf{RQ1.} Does \ours outperform state-of-the-art product summarization methods w.r.t. to different summarization metrics?

\item \textbf{RQ2.} Does each component in \ours contribute to its overall performance?

\item \textbf{RQ3.} Is \ours sensitive to the setting of task weights in Eq.~\ref{eq:eq-all}?

\item \textbf{RQ4.} Can \ours generate more coherent and descriptive summaries than baselines?

\end{itemize}

\begin{table*}[]
    \centering
    \resizebox{0.95\textwidth}{!}{
	\begin{tabular}{c|c|cccccccccc}
\hline
Category                                                                    & Method                & R-1            & R-2            & R-L            & B-1            & B-2            & B-3            & B-4            & S-B            & M              & BS             \\ \hline
\multirow{9}{*}{\begin{tabular}[c]{@{}c@{}}Home \\ Appliances\end{tabular}} & Lead                  & 22.15          & 9.46           & 19.59          & 26.08          & 19.30          & 14.49          & 11.17          & 10.58          & \textbf{35.95} & 63.34          \\
                                                                            & Seq2seq               & 26.88          & 7.98           & 20.91          & 39.32          & 25.53          & 17.01          & 11.48          & 9.66           & 21.44          & 65.26          \\
                                                                            & PG                    & 27.67          & 9.24           & 23.67          & 41.58          & 28.04          & 20.01          & 14.76          & 13.19          & 25.44          & 64.99          \\
                                                                            & MMPG                  & 32.88          & 11.88          & 21.96          & 36.20          & 19.07          & 10.40          & 5.54           & 2.90           & 16.47          & 65.13          \\
                                                                            & VG-BART (Multi-head)  & 32.10          & 11.72          & 25.15          & 48.34          & 34.25          & 24.92          & {\ul 18.59}    & {\ul 16.83}    & 29.58          & 67.73          \\
                                                                            & VG-BART (Dot-product) & 32.07          & 11.53          & {\ul 25.46}    & 48.25          & 33.87          & 24.45          & 18.13          & 16.47          & 29.52          & 67.72          \\
                                                                            & V2P                   & {\ul 34.47}    & {\ul 12.63}    & 25.09          & {\ul 52.23}    & {\ul 36.49}    & {\ul 25.30}    & 17.81          & 15.24          & 30.31          & {\ul 68.76}    \\
                                                                            & \ours                 & \textbf{34.91} & \textbf{13.20} & \textbf{26.44} & \textbf{54.22} & \textbf{38.65} & \textbf{28.26} & \textbf{21.22} & \textbf{19.66} & {\ul 32.32}    & \textbf{69.76} \\
                                                                            & Improvement           & 1.28\%         & 4.51\%         & 3.85\%         & 3.28\%         & 11.31\%        & 16.72\%        & 15.76\%        & 16.82\%        & -10.10\%       & 1.45\%         \\ \hline
\multirow{9}{*}{Clothing}                                                   & Lead                  & 19.84          & 7.13           & 17.77          & 16.78          & 10.76          & 7.18           & 5.10           & 4.01           & \textbf{28.41} & 60.61          \\
                                                                            & Seq2seq               & 31.33          & 9.88           & 23.02          & 35.71          & 21.84          & 14.29          & 9.54           & 7.43           & 23.46          & 68.22          \\
                                                                            & PG                    & 30.92          & 9.66           & 24.31          & 36.28          & 21.90          & 14.08          & 9.22           & 7.39           & 26.07          & 67.96          \\
                                                                            & MMPG                  & 30.73          & 10.29          & 21.25          & 28.32          & 11.69          & 5.15           & 0.57           & 2.24           & 14.47          & 66.07          \\
                                                                            & VG-BART (Multi-head)  & 31.32          & 10.80          & {\ul 23.49}    & 36.74          & 23.29          & 15.58          & {\ul 10.81}    & {\ul 8.69}     & 27.23          & 67.79          \\
                                                                            & VG-BART (Dot-product) & 30.18          & 10.40          & 22.82          & 35.30          & 22.37          & 15.09          & 10.59          & 8.36           & 26.32          & 66.81          \\
                                                                            & V2P                   & \textbf{35.05} & \textbf{11.98} & 22.62          & \textbf{42.41} & {\ul 25.04}    & {\ul 15.70}    & 9.16           & 6.55           & 27.96          & {\ul 68.40}    \\
                                                                            & \ours                 & {\ul 34.05}    & {\ul 11.52}    & \textbf{25.09} & {\ul 41.43}    & \textbf{26.05} & \textbf{17.17} & \textbf{11.62} & \textbf{9.26}  & {\ul 28.30}    & \textbf{69.57} \\
                                                                            & Improvement           & -2.85\%        & -3.84\%        & 10.92\%        & -2.31\%        & 4.03\%         & 9.36\%         & 7.49\%         & 6.56\%         & -0.39\%        & 1.71\%         \\ \hline
\multirow{9}{*}{\begin{tabular}[c]{@{}c@{}}Cases \& \\ Bags\end{tabular}}   & Lead                  & 20.15          & 7.32           & 18.26          & 17.93          & 11.71          & 7.89           & 5.60           & 4.56           & {\ul 29.23}    & 61.05          \\
                                                                            & Seq2seq               & 28.59          & 8.07           & 20.60          & 23.64          & 14.11          & 8.85           & 5.74           & 4.38           & 17.79          & {\ul 68.16}    \\
                                                                            & PG                    & 32.18          & 9.73           & 24.73          & 40.06          & 23.97          & 15.00          & 9.58           & 7.70           & 26.43          & 68.14          \\
                                                                            & MMPG                  & 32.69          & 11.78          & 22.27          & 30.39          & 12.46          & 5.26           & 2.14           & 0.49           & 14.93          & 65.44          \\
                                                                            & VG-BART (Multi-head)  & 30.76          & 10.46          & 24.88          & 38.16          & 24.31          & 16.22          & {\ul 11.30}    & {\ul 9.39}     & 25.81          & 67.11          \\
                                                                            & VG-BART (Dot-product) & 31.21          & 10.74          & {\ul 25.13}    & 38.61          & 24.71          & 16.27          & 11.09          & 9.10           & 26.39          & 67.81          \\
                                                                            & V2P                   & {\ul 34.65}    & {\ul 11.89}    & 24.53          & {\ul 43.88}    & {\ul 27.32}    & {\ul 16.64}    & 10.17          & 7.40           & 28.25          & 67.74          \\
                                                                            & \ours                 & \textbf{35.08} & \textbf{12.08} & \textbf{25.60} & \textbf{44.35} & \textbf{28.20} & \textbf{18.45} & \textbf{12.45} & \textbf{10.57} & \textbf{29.37} & \textbf{68.95} \\
                                                                            & Improvement           & 1.24\%         & 1.60\%         & 1.87\%         & 1.07\%         & 3.22\%         & 10.88\%        & 10.18\%        & 12.57\%        & 0.48\%         & 1.16\%         \\ \hline
\end{tabular}%
    }
    \caption{Performance of all methods on three product categories. The best results are shown in bold and the second-best results are underlined. The percentages of the improvement are obtained by comparing \ours with the best baseline. ROUGE, BLEU, SENTENCE-BLEU, METEOR and BERTScore are denoted by R, B, S-B, M and BS, respectively.}
    \label{per_com}
    \end{table*}

\subsection{Experimental Settings}

\vspace{5pt}
\noindent\textbf{Dataset.} 
We use the CEPSUM dataset\footnote{\url{https://github.com/hrlinlp/cepsum}}~\cite{LiYXWHZ20}, which is collected from an e-commerce platform in China.
It includes around 1.4 million products covering three categories: Home Appliances, Clothing, and Cases \& Bags. 
Each product in CEPSUM contains a long product description, a product title, a product image, and a high-quality summary written by humans. 
Tab.~\ref{data_static} provides the data statistics.
 
For the multi-modal product attribute prediction task, we construct the pre-defined attribute vocabulary following V2P~\cite{SongJLZCN22}. 
Specifically, we use Jieba\footnote{\url{https://github.com/fxsjy/jieba}} to tokenize the dataset and then perform the part-of-speech tagging. 
The attribute vocabulary for each category consists of adjectives and nouns with more than one Chinese character and have appeared in more than a pre-defined number of products' summaries. According to the scale of different product categories, the pre-defined number threshold for Home Appliance, Clothing, and Cases \& Bags categories are set to $5,000$, $10,000$, and $1,000$, respectively.

\vspace{5pt}
\noindent\textbf{Implementation Details.} 
We use the pre-trained bart-base-chinese\footnote{\url{https://huggingface.co/fnlp/bart-base-chinese}} to construct the text encoder, which has six layers.
For the image encoder, we stack 4 layers (i.e., $H$ in Sec.~\ref{sec:ie}) with $8$ attention heads and the hidden dimensionality is $2,048$. 
The parameters of the image encoder are initialized randomly. 
The batch size $B$ of Eq.~\ref{eq:cl} is $16$.
The dropout rate is $0.1$. 
The maximum length $L$ of the text sequence is set to $400$, and the retained number $M$ of regions per image is set to $36$. 
We conduct a grid search for task weights $\lambda_{1}$, $\lambda_{2}$ and $\lambda_{3}$ in Eq.~\ref{eq:eq-all}. 
The default $\lambda_{1}$, $\lambda_{2}$ and $\lambda_{3}$ are set to $0.8$, $0.05$ and $0.3$, respectively.
We also report the impact of performance when using different task weights in Sec.~\ref{sec:sa}.
The models are optimized using Adam~\cite{KingmaB14} and a cosine learning rate schedule with the initial learning rate of $3e^{-5}$.  
Notably, we report the average of three runs with different random seeds on the testing set as experimental results in this paper.

\begin{figure*}[t]  
    \centering  
    \includegraphics[width=0.92\textwidth]{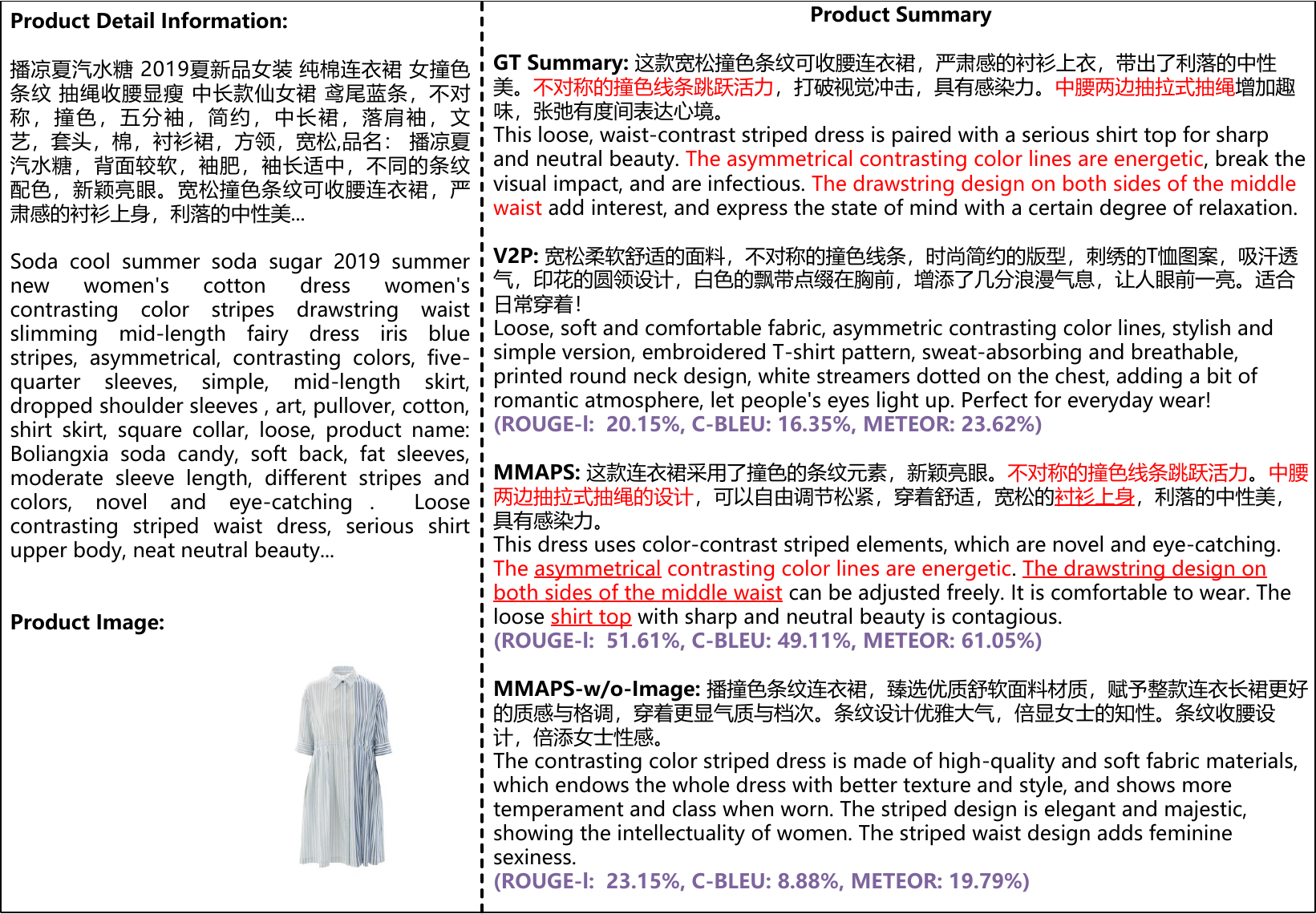}  
    \caption{A comparison between the product summaries generated by \ours and V2P. GT indicates ground truth. 
    The English texts are translated from the corresponding Chinese texts. The same or semantically similar descriptions are highlighted in red. The product appearances manifested in the image are underlined.}
    \label{fig:casestudy}  
\end{figure*}

\vspace{5pt}
\noindent\textbf{Evaluation Metric.}  
We use ROUGE (1, 2, L)~\cite{lin2004rouge}, BLEU (1, 2, 3, 4)~\cite{apineniRWZ02}, S-BLEU~\cite{NLTK}, METEOR~\cite{BanerjeeL05} and BERTScore~\cite{ZhangKWWA20} as the evaluation metrics.
BLEU measures the micro-average precision.
We also adopt S-BLEU that averages the sentence-level BLEU-4 scores (i.e., macro-average precision).
Instead of exact matching, BERTScore measures the relevance between generated summaries and ground truth in the semantic space of BERT~\cite{DevlinCLT19}. 
Thus, it correlates with human judgments.

\vspace{5pt}
\noindent\textbf{Baselines.}
We adopt the following methods, which are prevalently used in the experiments of existing product summarization works~\cite{LiYXWHZ20,SongJLZCN22}, as baselines in our experiments: 
\begin{itemize}[leftmargin=10pt,topsep=1pt,itemsep=0.3pt]
\item \textbf{Lead~\cite{LiYXWHZ20,SongJLZCN22}} directly extracts the first 80 characters of the long product descriptions as product summaries.

\item \textbf{Seq2seq~\cite{abs-1807-08000}} is the standard neural architecture used for text generation. It takes the long product descriptions as input and outputs corresponding product summaries.

\item \textbf{Pointer-Generator~\cite{SeeLM17}} is a hybrid method consisting of the pointer network and the Seq2seq architecture. The pointer network helps reproduce input in the generated summaries by copying words from the long product descriptions.

\item \textbf{MMPG\footnote{\url{https://github.com/hrlinlp/cepsum}}~\cite{LiYXWHZ20}} is a multi-modal dual-encoder pointer-generator network for product summarization, where the convolutional neural networks are used to encode product images.

\item \textbf{VG-BART (Dot-product) and VG-BART (Multi-head)\footnote{\url{https://github.com/HLTCHKUST/VG-GPLMs}}~\cite{YuDLF21}} adopt BART as the backbone for summarization, and use dot-product based fusion and multi-head based fusion to inject visual information, respectively.

\item \textbf{V2P\footnote{\url{https://xuemengsong.github.io/V2P_Code.rar}}~\cite{SongJLZCN22}} adopts BART as the backbone. It enhances product summarization by predicting product attributes and using attribute prompts extracted from product images.

\end{itemize}

\subsection{Overall Performance (RQ1)}
\label{sec:op}

We report the results in Tab.~\ref{per_com}.
From the results, we can observe that: 
\begin{enumerate}[leftmargin=14pt,topsep=1pt,itemsep=0.3pt]

\item \ours achieves the best results in most cases, especially outperforming other methods by a large margin on BLEU metrics. VG-BART, V2P and \ours all use the same PLM (BART) as the backbone, but \ours performs much better than others in terms of most metrics. For example, our method exceeds the best baseline by $15.76\%$ and $16.82\%$ for BLEU-4 and S-BLEU on Home Appliances, respectively. The results indicate that \ours consistently generates high-quality product summaries that are closer to human-written summaries.

\item In a few cases, baselines outperform \ours. However, they do not show robust performance as \ours. 
For example, Lead shows a superior performance on the METEOR metric, which considers the number of chunks. 
Moreover, as explained by the authors of V2P~\cite{SongJLZCN22}, V2P obtains better scores on Rouge-1 and Rouge-2 on the Clothing category (the performance of \ours is 2.85\% and 3.84\% worse than V2P on Rouge-1 and Rouge-2) since human-written summaries in the Clothing category are more likely to contain the vision-related attributes.
However, compared to \ours, both Lead and V2P do not show consistently good performance on all metrics across the three categories.

\item We also find that V2P obtains better results than other baselines on unigram and bigram metrics (R-1, R-2, B-1 and B-2) while it is inferior to VG-BART and \ours on trigram, 4-gram and sentence-level metrics (R-L, B-3, B-4 and S-B). The results indicate that, compared to VG-BART and \ours, V2P cannot generate summaries containing adjectives and nouns with relatively more words (e.g., 4-gram text), making the generated summaries of V2P less coherent than those generated by \ours. This is also observed in the case study reported in Sec.~\ref{sec:cs}.

\end{enumerate}

\begin{figure*}[htbp]
    \includegraphics[width=\textwidth]{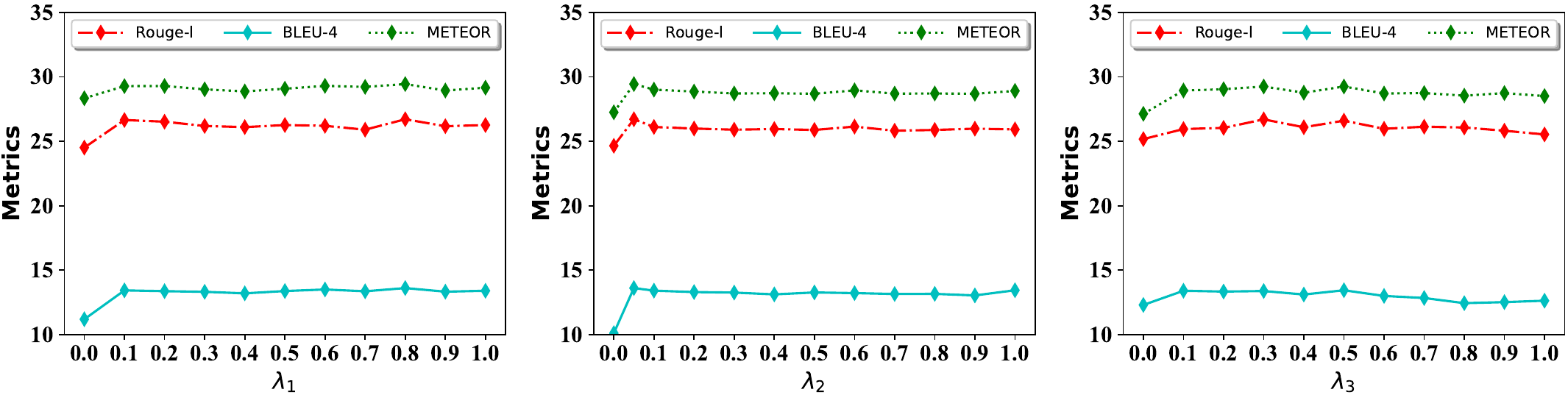}
    \caption{Sensitivity analysis of task weights on Cases \& Bags.}
    \label{fig:sen_ana}
\end{figure*}

\subsection{Ablation Study (RQ2)} 

\begin{table*}[t]
    \centering
    \resizebox{0.97\linewidth}{!}{
    \begin{tabular}{c|cccccccccc}
    \hline
    Model         & R-1            & R-2            & R-L            & B-1            & B-2            & B-3            & B-4            & S-B            & M              & BS             \\ \hline
    \ours         & \textbf{35.08} & \textbf{12.08} & \textbf{25.60} & 44.35          & \textbf{28.20} & \textbf{18.45} & \textbf{12.45} & \textbf{10.57} & \textbf{29.37} & \textbf{69.76} \\
    \ours-w/o-MRM & 34.44          & 11.35          & 24.51          & 44.20          & 27.58          & 17.44          & 11.19          & 8.79           & 28.34          & 69.74          \\
    \ours-w/o-CMM & 33.88          & 10.32          & 24.65          & 43.10          & 25.89          & 15.95          & 10.11          & 7.96           & 27.25          & 68.63          \\
    \ours-w/o-FMM & 34.05          & 11.76          & 24.81          & \textbf{44.55} & 27.88          & 18.17          & 12.30          & 9.84           & 27.13          & 69.40           \\ \hline
    \end{tabular}
    }
    \caption{The results of ablation studies on Cases \& Bags. The best results are shown in bold.}
    \label{ablation}
\end{table*}

To investigate the contribution of each component in \ours, we design the following variants of \ours for the ablation study:
\begin{itemize}[leftmargin=10pt,topsep=1pt,itemsep=0.3pt]

\item \textbf{\ours-w/o-MRM}: To verify the importance of 
the masked region modeling task, we remove it by setting $\lambda_{1} = 0$ in Eq.~\ref{eq:eq-all}.

\item \textbf{\ours-w/o-CMM}: To show the effect of coarse-grained multi-modal modeling, we take this task out by setting $\lambda_{2} = 0$ in Eq.~\ref{eq:eq-all}.

\item \textbf{\ours-w/o-FMM}: To show the necessity of fine-grained multi-modal modeling, 
we exclude it by setting $\lambda_{3}$ to zero in Eq.~\ref{eq:eq-all}.

\end{itemize}

Due to page limit, we only show the ablation study results of the above methods over the Cases \& Bags in Tab.~\ref{ablation}. 
We have the following findings:
\begin{enumerate}[leftmargin=14pt,topsep=1pt,itemsep=0.3pt]
\item The mask region modeling task improves the quality of the generated summaries on all metrics, demonstrating that predicting the class of the masked regions is helpful to the image encoder and it enhances product summarization.

\item Introducing the coarse-grained multi-modal modeling exhibits notable benefits, showing that capturing the semantic correlation between two modalities at the sequence level can facilitate generating high-quality product summaries.

\item The two fine-grained multi-modal modeling tasks show a positive impact on the model performance, suggesting that refining the fine-grained token-level and region-level representations and capturing their mutual influence are beneficial to product summary generation.
\end{enumerate}

\subsection{Impact of Task Weights (RQ3)}
\label{sec:sa}

We inspect the sensitivity of \ours to task weights in Eq.~\ref{eq:eq-all} over the category Cases \& Bags. 
We vary the values from 0 to 1 at the step of 0.1. 
From Fig.~\ref{fig:sen_ana} we can see that \ours performs worst when setting one of the three task weights to zero.
This confirms the importance of the masked region modeling task, the coarse-grained multi-modal modeling and the fine-grained multi-modal modeling to the overall performance of \ours.
Besides, we observe that \ours performs relatively stably when the three task weights are non-zero. 
This implies that \ours is not sensitive to different task weights as long as they are non-zero.

\subsection{Case Study (RQ4)} 
\label{sec:cs}

Fig.~\ref{fig:casestudy} provides a case study for better understanding the difference between \ours and V2P, which is the best baseline in most cases.
From Fig.~\ref{fig:casestudy}, we have the following observations:
\begin{enumerate}[leftmargin=14pt,topsep=1pt,itemsep=0.3pt] 

\item V2P cannot generate grammarly corrected summaries and it produces many short phrases containing only product attributes. This phenomenon explains why V2P shows good performance on BLEU-1 in Tab.~\ref{per_com}, but performs poorly when tested using n-gram based metrics with a large $n$ (e.g., BLEU-3, BLEU-4 and S-BLEU). Differently, \ours can generate more coherent and descriptive product summaries containing rich product attribute information described by n-grams with a large $n$. For example, 
the generated summary of \ours contains several n-grams with a relatively large $n$, e.g., ``The asymmetrical contrasting color lines are energetic'' and ``the drawstring design on both sides of the middle waist''. And these n-grams are also contained in the human-written summary.
This observation shows that \ours can capture the product attributes which are better described using n-grams with large n values. Therefore, \ours can generate more coherent and readable summaries as discussed in Sec.~\ref{sec:op}.

\item From the generated summaries, we can see that \ours is able to model the product appearances manifested in the image (i.e., the underlined parts: ``asymmetrical'', ``shirt top'' and ``The drawstring design on both sides of the middle waist''), while \ours-w/o-Image cannot. Hence, the image modality indeeds enhances the product summarization by providing additional signal to guide the model to generate the attractive summaries.

\end{enumerate}

\section{Conclusion}

In this paper, we propose \ours for multi-modal product summarization and it is able to model product attributes and produce coherent and descriptive summaries simultaneously. 
Our experiments show that \ours exceeds state-of-the-art product summarization methods.
In the future, we plan to incorporate cross-grained contrast learning, i.e., the contrast between coarse-grained representations and fine-grained representations.
We will also consider generating personalized product summarization based on user preferences and affinity towards different product characteristics.

\section*{Acknowledgments}
This work was partially supported by National Science and Technology Major Project (No. 2022ZD0118201) and National Natural Science Foundation of China (No. 62002303, 42171456).

\clearpage

\section*{References}
\label{sec:reference}
\vspace{-20pt}

\balance
\bibliographystyle{lrec-coling2024-natbib}
\bibliography{ref}

\end{document}